\RequirePackage{fix-cm}
\documentclass[twocolumn,epjc3]{svjour3}  
\smartqed  
\RequirePackage{graphicx}

\usepackage{amsmath, amsfonts,amssymb}
\usepackage{xparse}
\usepackage{mathrsfs}
\usepackage{breqn}
\usepackage{braket}

\newcommand{\Ag}{\mathcal{A}}

\newcommand*{\de}[1]{\mathop{\mathrm{d}#1}\nolimits}

\NewDocumentCommand\MyAc{ m }{#1}
\DeclareDocumentCommand{\ct}{ t. t, t- t' s s m m m }{
  \RenewDocumentCommand\MyAc{ m }{##1}
  \IfBooleanT{#1}{\RenewDocumentCommand\MyAc{ m }{ \mathring{##1} } }
  \IfBooleanT{#2}{\RenewDocumentCommand\MyAc{ m }{ \tilde{##1} } }
  \IfBooleanT{#3}{\RenewDocumentCommand\MyAc{ m }{ \bar{##1} } }
  \IfBooleanT{#4}{\RenewDocumentCommand\MyAc{ m }{ {##1}' } }
  \IfBooleanTF{#5}
  { \IfBooleanTF{#6} { \hat{\MyAc{\Gamma}}_{{#7}}{}^{\hat{#8}}{}_{\hat{#9}} }{ \hat{\MyAc{\Gamma}}_{{#7}}{}^{{#8}}{}_{{#9}} } }
  { \MyAc{\Gamma}_{{#7}}{}^{{#8}}{}_{{#9}} } }
\DeclareDocumentCommand{\ri}{ t. t, t- s s m m m }{
  \RenewDocumentCommand\MyAc{ m }{##1}
  \IfBooleanT{#1}{\RenewDocumentCommand\MyAc{ m }{ \mathring{##1} } }
  \IfBooleanT{#2}{\RenewDocumentCommand\MyAc{ m }{ \tilde{##1} } }
  \IfBooleanT{#3}{\RenewDocumentCommand\MyAc{ m }{ \bar{##1} } }
  \IfBooleanTF{#4}
  { \IfBooleanTF{#5} { \hat{\MyAc{\mathcal{R}}}_{{#6}}{}^{\hat{#7}}{}_{\hat{#8}} }{ \hat{\MyAc{\mathcal{R}}}_{{#6}}{}^{{#7}}{}_{{#8}} } }
  { \MyAc{\mathcal{R}}_{{#6}}{}^{{#7}}{}_{{#8}} } }
\DeclareDocumentCommand{\bt}{ t. t, t- s s m m m }{
  \RenewDocumentCommand\MyAc{ m }{##1}
  \IfBooleanT{#1}{\RenewDocumentCommand\MyAc{ m }{ \mathring{##1} } }
  \IfBooleanT{#2}{\RenewDocumentCommand\MyAc{ m }{ \tilde{##1} } }
  \IfBooleanT{#3}{\RenewDocumentCommand\MyAc{ m }{ \bar{##1} } }
  \IfBooleanTF{#4}
  { \IfBooleanTF{#5} { \hat{\MyAc{\mathcal{B}}}_{{#6}}{}^{\hat{#7}}{}_{\hat{#8}} }{ \hat{\MyAc{\mathcal{B}}}_{{#6}}{}^{{#7}}{}_{{#8}} } }
  { \MyAc{\mathcal{B}}_{{#6}}{}^{{#7}}{}_{{#8}} } }

\newcommand\UTFSM{Departamento de F\'isica, Universidad T\'{e}cnica Federico Santa Mar\'\i a\\ Casilla 110-V, Valpara\'iso, Chile}

\newcommand\CCTVal{Centro Cient\'ifico Tecnol\'ogico de Valpara\'iso\\ Casilla 110-V, Valpara\'\i so, Chile}
\newcommand{\UdelaR}{Instituto de F\'isica, Facultad de Ciencias\\Igu\'a 4225, esq. Mataojo, 11400 Montevideo, Uruguay.}
\newcommand{\IFIC}{Departamento de F\'isica Te\'orica and IFIC, Centro Mixto Universidad de Valencia - CSIC\\ Universidad de Valencia, Burjassot-46100, Valencia, Spain.}

\RequirePackage[numbers,sort&compress]{natbib}
\RequirePackage[colorlinks,citecolor=blue,urlcolor=blue,linkcolor=blue]{hyperref}
\journalname{Eur. Phys. J. C}
\begin{document}

\title{Inflationary scenarios in an effective polynomial affine model of gravity\thanksref{t1}
}

\titlerunning{Inflation in polynomial affine gravity}        

\author{Oscar Castillo-Felisola\thanksref{e1,addr1,addr2}
        \and
        Bastian Grez\thanksref{addr1}
        \and
        Jos\'e Perdiguero\thanksref{e2,addr3}
        \and
        Aureliano Skirzewski\thanksref{addr4}
}

\thankstext{t1}{This work has been fund by ANID PIA/APOYO AFB220004 (Chile) and FONDECYT Grant 1230110 (Chile)}
\thankstext{e1}{e-mail: o.castillo.felisola:at:proton.me}
\thankstext{e2}{e-mail: jose.perdiguerog:at:gmail.com}


\institute{\UTFSM \label{addr1}
           \and
           \CCTVal \label{addr2}
           \and
           \IFIC \label{addr3}
           \and
           \UdelaR \label{addr4}
}

\date{Received: date / Accepted: date}

\maketitle

\begin{abstract}
  In this paper we inquire inflationary scenarios built on a simplified version of the polynomial affine model of gravity. Given the absence of a metric tensor in the formulation of the model, we build a \emph{kinetic term} contracting the derivatives of scalar field with the most general \((2,0)\)-tensor density build using the affine connection, and introduce a self-interacting potential via a scaling of the volume form. We analyse the cosmological solutions derived from this setup.
\end{abstract}

\section{Introduction}
\label{sec:intro}
Modern cosmological models propose the existence of a period of exponential expansion of the (very) early Universe, known as inflation, in order to explain its current large-scale structure. Inflationary models have become essential in our understanding of the evolution of the Universe, as they provide solutions to challenging problem occurring in the standard Big Bang cosmology, for example the flatness, horizon and relic density problems \cite{liddle00_cosmol,lyth09,weinberg08_cosmol,baumann22_cosmol}.

Typically, inflationary models involve scalar fields coupled with gravity, and hence their dynamics can be studied in various gravitational theories. The laxity in the choice of gravitational model and its interaction with the scalar sector, turns the theoretical modelling of inflation into a fertile soil that raises many inflationary scenarios \cite{martin13_encyc_inflat}. However, the inflationary dynamics is customarily analysed in metrical theories of gravity, such as General Relativity.

The purpose of this work is to test a simple inflationary scenario occurring when a scalar field is coupled to the polynomial affine model of gravity \cite{castillo-felisola15_polyn_model_purel_affin_gravit,castillo-felisola18_einst_gravit_from_polyn_affin_model}. An interesting feature of the polynomial affine model of gravity is that the absence of a metric tensor field in its formulation constraint the number of terms that can be added to the action functional, endowing it with a sort of \emph{rigidity} which is not present in other models of gravity, including other affine gravities \cite{eddington23,einstein23_zur_affin_feldt,schroedinger50_space,kijowski78_new_variat_princ_gener_relat,kijowski07_univer_affin_formul_gener_relat,krasnov07_non_metric_gravit,krasnov11_pure_connec_action_princ_gener_relat,poplawski07_nonsy_purel_affin,poplawski14_affin_theor_gravit,demir14_rieman}.

Naturally, one could question the relevance of analysing affine models of gravity provided that the structure of the observed Universe possesses a metric that allows us to measure distances. However, it might be argue that either: (i) the metrical description might be originated from the dynamics of an affine Universe; (ii) the metric structure, although fundamental for determining the geometrical properties of the spacetime, plays no role in the formulation of the gravitational sector;\footnote{Curiously, in metrical models of gravity, the metric plays a double role, as mediator of the gravitational interactions and instrument for measuring distances (it is also the responsible of providing a notion of parallelism). The exclusion of the metric in the formulation of the model takes out the responsibility of mediating the interaction (and defining parallelism), leaving it the sole role of a \emph{compass}.} or (iii) alternative formulations of gravity, even if equivalent to standard General Relativity (as in the case of the affine model considered by Eddington and Einstein  \cite{eddington23,einstein23_zur_affin_feldt}, or teleparallel gravity \cite{aldrovandi13_telep_gravit}), might be inequivalent if one attempts to extend them \cite{krasnov20_formul_gener_relat}.

Unlike in metric theories, there is no standard way to couple matter to affine gravity, even if the \emph{minimal coupling} procedure could be defined, the absence of the metric tensor prevents the introduction of a kinetic term, i.e. \(g^{\mu\nu} \partial_{\mu} \phi \partial_{\nu} \phi\). In non-polynomial affine models, as for example the one Eddington--Einstein, the kinetic tensor \(\partial_{\mu} \phi \partial_{\nu} \phi\) can be added to the Ricci tensor originating their interaction once the determinant is calculated \cite{kijowski07_univer_affin_formul_gener_relat}. Using this approach, some inflationary models from affine gravity were studied in Refs. \cite{azri17_affin_inflat,azri18_induc_affin_inflat,azri18_cosmol_implic_affin_gravit}.

Just like a point of comparison, it is curious to notice that in metric and metric-affine gravitational models, the use of the inverse metric as part of the formulation (i.e. it appears explicitly in the action functional) allows us to include infinitely many terms to the action, unless there are additional arguments to cut the number of term, such as the preservation of certain superficial degree of freedom \cite{weinberg95_quant_theor_field}. Horndenski found the most general tensor-scalar functional that leads to second order field equations \cite{horndeski74_secon_order_scalar_tensor}, and his model has gain importance in the formulation of different models of inflation and dark energy \cite{clifton12_modif_gravit_cosmol}. Oppositely, the rigidity of the polynomial affine gravity constraints the possible couplings between the gravitational and scalar sector. 

The aim of this article is to examine the simplest inflationary models obtained by coupling the kinetic tensor of a scalar field with the polynomial affine model of gravity. For the sake of simplicity, we restrict ourselves to the torsion-free sector of the model, which in the case of pure gravity yields field equations that generalise those of General Relativity. The article is organised as follows: In section \ref{sec:setup} we recapitulate what polynomial affine gravity is, highlighting and presenting our simplified setup; Section \ref{sec:solutions} begins by re-examining the known solutions to the purely gravitational system in section \ref{sec:purely-pag}, to move later to analyse the cases coupled with the scalar field in section \ref{sec:coupled-pag}; Issues regarding our solutions are discussed in section \ref{sec:discussion}, along with our concluding remarks. For the sake of completeness, we prove in appendix \ref{app:parallel-tensor} that a symmetric \(\binom{0}{2}\)-tensor compatible with the symmetries of the cosmological principle is parallel if and only if the connection is Riemannian.

\section{The simplified model}
\label{sec:setup}
The polynomial affine model of gravity is a model built up solely with the irreducible components of an affine connection \(\ct,{\mu}{\lambda}{\rho}\), denoted in here on by \(\ct{\mu}{\lambda}{\rho}\), \(\Ag_{\mu}\) and \(\bt{\mu}{\lambda}{\rho}\) defined by
\begin{equation}
  \ct,{\mu}{\lambda}{\rho} = \ct{\mu}{\lambda}{\rho} + \Ag_{[\mu} \delta^{\lambda}_{\rho]} + \bt{\mu}{\lambda}{\rho}.
  \label{eq:conn}
\end{equation}
Note that \(\ct{\mu}{\lambda}{\rho} = \ct,{(\mu}{\lambda}{\rho)}\) is the symmetric part of affine connection, \(\Ag_{\mu}\) is related to the trace of the torsion \(\ct{\mu}{\lambda}{\lambda}\) and \(\bt{\mu}{\lambda}{\rho}\) corresponds to the trace-free part of the connection.

The action of the polynomial affine model of gravity in four-dimensions \cite{castillo-felisola15_polyn_model_purel_affin_gravit} (see Ref. \cite{castillo-felisola20_emerg_metric_geodes_analy_cosmol} for the new parametrisation) does not require the use of a metric tensor field to be built up, but needs a volume form \(\de{V}^{\alpha \beta \gamma \delta}\). The most general action (up to topological and boundary terms) is given by
\begin{dmath}[style={\small}]
  \label{eq:new-action}
  S  = \int \de{V}^{\alpha \beta \gamma \delta} \bigg[ B_1
  \ri{\mu\nu}{\mu}{\rho} \bt{\alpha}{\nu}{\beta}
  \bt{\gamma}{\rho}{\delta} + B_2 \ri{\alpha\beta}{\mu}{\rho}
  \bt{\gamma}{\nu}{\delta} \bt{\mu}{\rho}{\nu} + B_3
  \ri{\mu\nu}{\mu}{\alpha} \bt{\beta}{\nu}{\gamma} \Ag_\delta + B_4
  \ri{\alpha\beta}{\sigma}{\rho} \bt{\gamma}{\rho}{\delta} \Ag_\sigma
  + B_5 \ri{\alpha\beta}{\rho}{\rho}
  \bt{\gamma}{\sigma}{\delta} \Ag_\sigma + C_1
  \ri{\mu\alpha}{\mu}{\nu} \nabla_\beta \bt{\gamma}{\nu}{\delta} +
  C_2 \ri{\alpha\beta}{\rho}{\rho} \nabla_\sigma
  \bt{\gamma}{\sigma}{\delta} + D_1 \bt{\nu}{\mu}{\lambda}
  \bt{\mu}{\nu}{\alpha} \nabla_\beta \bt{\gamma}{\lambda}{\delta}
  + D_2 \bt{\alpha}{\mu}{\beta} \bt{\mu}{\lambda}{\nu}
  \nabla_\lambda \bt{\gamma}{\nu}{\delta} + D_3 \bt{\alpha}{\mu}{\nu}
  \bt{\beta}{\lambda}{\gamma} \nabla_\lambda \bt{\mu}{\nu}{\delta} +
  D_4 \bt{\alpha}{\lambda}{\beta} \bt{\gamma}{\sigma}{\delta}
  \nabla_\lambda \Ag_\sigma + D_5 \bt{\alpha}{\lambda}{\beta}
  \Ag_\sigma \nabla_\lambda \bt{\gamma}{\sigma}{\delta}
  + D_6 \bt{\alpha}{\lambda}{\beta} \Ag_\gamma \nabla_\lambda
  \Ag_\delta + D_7 \bt{\alpha}{\lambda}{\beta} \Ag_\lambda
  \nabla_\gamma \Ag_\delta + E_1 \nabla_\rho \bt{\alpha}{\rho}{\beta}
  \nabla_\sigma \bt{\gamma}{\sigma}{\delta} + E_2 \nabla_\rho
  \bt{\alpha}{\rho}{\beta} \nabla_\gamma \Ag_{\delta}
  + F_1 \bt{\alpha}{\mu}{\beta} \bt{\gamma}{\sigma}{\delta}
  \bt{\mu}{\lambda}{\rho} \bt{\sigma}{\rho}{\lambda} + F_2
  \bt{\alpha}{\mu}{\beta} \bt{\gamma}{\nu}{\lambda}
  \bt{\delta}{\lambda}{\rho} \bt{\mu}{\rho}{\nu} + F_3
  \bt{\nu}{\mu}{\lambda} \bt{\mu}{\nu}{\alpha}
  \bt{\beta}{\lambda}{\gamma} \Ag_\delta + F_4
  \bt{\alpha}{\mu}{\beta} \bt{\gamma}{\nu}{\delta} \Ag_\mu \Ag_\nu
  \bigg].
\end{dmath}
In the action, the covariant derivation and curvature tensors are defined with respect to the symmetrised connection, i.e. \(\nabla = \nabla^{(\Gamma)}\) and \(\mathcal{R} = \mathcal{R}(\Gamma)\). The \emph{uniqueness} of the action relies in a sort of \emph{dimensional analysis} provided by the indices structure of the fields, and the symmetries of the geometric quantities \cite{castillo-felisola18_einst_gravit_from_polyn_affin_model}.

Some of the features of the action in Eq. \eqref{eq:new-action} are: (i) Its rigidity, since contains all possible combinations of the fields and their derivatives; (ii) All the coupling constants are dimensionless, which might be a sign of \emph{conformal} invariance, and also ensure that the model is power-counting renormalisable; (iii) The field equations are second order differential equations, and the Einstein spaces are a subset of their solutions; (iv) The supporting symmetry group is the group of diffeomorphisms, desirable for the background independence of the model; (v) Even though there is no fundamental metric, it is possible to obtain \emph{emergent} (connection-descendent) metric tensors; (vi) The cosmological constant appears in the solutions as an integration constant, changing the paradigm concerning its nature;\footnote{This is similar to what happens in the unimodular model of gravity \cite{ng91_unimod_theor_gravit_cosmol_const,jirousek23_unimod_approac_to_cosmol}.} (vii) The model can be extended to be coupled with a scalar field, and the field equations are \emph{equivalent} to those of General Relativity interacting with a massless scalar field.

The key issue to couple a scalar field with the polynomial affine model of gravity is the possibility of defining a symmetric \(\binom{2}{0}\)-tensor density,\footnote{This tensor density can be obtained using the same analysis of the indices structure that allows to determine the action. For details see Ref. \cite{castillo-felisola18_einst_gravit_from_polyn_affin_model}.} by
\begin{dmath}
  \mathfrak{g}^{\mu\nu} = \alpha \, \nabla_{\lambda} \bt{\rho}{(\mu}{\sigma} \de{V}^{\nu)\lambda\rho\sigma}
    + \beta \, \Ag_{\lambda} \bt{\rho}{(\mu}{\sigma} \de{V}^{\nu)\lambda\rho\sigma}
    + \gamma \, \bt{\kappa}{\mu}{\lambda} \bt{\rho}{\nu}{\sigma} \de{V}^{\kappa\lambda\rho\sigma},
  \label{eq:inv_metr_dens}
\end{dmath}
which plays the role of \emph{inverse metric density}. In the above equation \(\alpha\), \(\beta\) and \(\gamma\) arbitrary constants, which can be coupled to the partial derivatives of the scalar field to provide a \emph{kinetic term}, as follows,
\begin{equation}
  S_{\phi} = - \int \mathfrak{g}^{\mu\nu} \partial_{\mu} \phi \partial_{\nu} \phi.
  \label{eq:scalar_kin_term}
\end{equation}

The complete action of the polynomial affine model of gravity coupled with a scalar field is given by the sum of the actions in Eqs. \eqref{eq:new-action} and \eqref{eq:scalar_kin_term}. However, for the purpose of this work we shall consider a simplified model, which is inspired in the effective model in the torsion-free sector. In this sector only the terms with coupling constants \(C_1\) and \(C_2\) contribute to the field equations, but in the cosmological scenario the model is restricted further and the sole component of the action \eqref{eq:new-action}  contributing to the field equations is the one with coupling \(C_1\). Similarly, the only term from the action \eqref{eq:scalar_kin_term} contributing to the field equations is the one with coupling \(\alpha\), since it is linear in the irreducible components of the torsion.

Therefore, we shall consider the simplified effective action,
\begin{equation}
  S = \int \de{V}^{\alpha \beta \gamma \delta} \left( \ri{\mu\alpha}{\mu}{\nu} - C \, \partial_{\alpha} \phi \partial_{\nu} \phi \right) \nabla_{\beta} \bt{\gamma}{\nu}{\delta},
  \label{eq:simple_model}
\end{equation}
where the coupling constant \(C\) is defined by the rate \(\frac{\alpha}{C_1}\).

The field equations derived from the action in Eq. \eqref{eq:simple_model} are
\begin{equation}
  \nabla_{\mu} \left( \left( C \, \partial_\alpha \phi \partial_\lambda \phi - \mathcal{R}_{\alpha\lambda} \right) \mathrm{d}V^{\mu\nu\rho\alpha} + \frac{2}{3} \mathcal{R}_{\alpha\theta} \delta^{[\nu}_{\lambda }\mathrm{d}V^{\rho]\alpha\mu\theta} \right)  = 0.
  \label{eq:simple_model_feqs}
\end{equation}
Interestingly, the above is the only nontrivial field equation of the system, and it is obtained optimising the effective action functional with respect to the \(\mathcal{B}\)-field. \cite{bekenstein15_is_princ_least_action}.
Notice that in the absence of a fundamental metric, the volume form is no longer associated with the metric (like in General Relativity) and hence it is not parallel in general. It would be useful for the remaining of the article to discuss further about the volume form.

On a four-dimensional manifold, \(\mathcal{M}\), a volume form is an everywhere nonvanishing four-form, and hence volume forms are defined up to a nonvanishing scalar, i.e. if \(\de{V}\) is a volume form the quantity \(\de{V}^{\prime}\) defined by \(\de{V}^{\prime\alpha\beta\gamma\delta} = J(x) \de{V}^{\alpha\beta\gamma\delta}\) is another volume form if \(J(x)\) is nowhere vanishing. A \emph{special} volume form is given by the epsilon density, \(\mathfrak{E}^{\alpha\beta\gamma\delta}\) with \(\mathfrak{E}^{0123} = 1\), because it is a density of weight \(+1\) whose definition is valid in every coordinate system \cite{schouten89_tensor,schouten13_ricci}.

Since every volume form \(\mathrm{d}V\) differs from the canonical volume form, \(\mathfrak{E}\), by a point-wise scaling, a mechanism to generate inflation on (a modified version of the Einstein--Eddington model of) affine gravity has been proposed by Azri and collaborators \cite{azri17_affin_inflat,azri18_induc_affin_inflat,azri18_cosmol_implic_affin_gravit}, where the scalar potential comes as a \emph{multiplicative} factor to the volume form. Geometrically, the scalar potential can be interpreted as coming from a variable volume form, which depends on the scalar field, i.e.
\begin{equation}
  \mathrm{d}V^{\alpha\beta\gamma\delta} = \frac{\mathfrak{E}^{\alpha\beta\gamma\delta}}{V(\phi)}.
  \label{eq:scaled-volume}
\end{equation}
This \emph{scaling} is analogous to the substitution of the metric tensor field from \(g_{\mu\nu}(x) \to g_{\mu\nu}(\phi)\), that allows us to get a non-linear \(\sigma\)-model from the standard kinetic term of a scalar field \(\phi\).

Since both \(\mathfrak{E}^{\alpha\beta\gamma\delta}\) and \(\mathfrak{e}_{\alpha\beta\gamma\delta}\) (the last one is the canonical skew-symmetric tensor densities of weight \(-1\)) are covariantly constant (aka \emph{parallel}), \(\nabla \mathfrak{E} = \nabla \mathfrak{e} = 0\), and satisfy the relation
\begin{equation}
  \mathfrak{E}^{\alpha\beta\gamma\delta} \mathfrak{e}_{\mu\nu\lambda\rho} = 4! \delta^{\alpha}_{[\mu} \delta^{\beta}_{\nu} \delta^{\gamma}_{\lambda} \delta^{\delta}_{\rho]},
  \label{eq:epsilon-to-delta}
\end{equation}
we can use them to define \emph{quasi-Hodge dualities} \cite{vaz16_introd_cliff_algeb_spinor}, which allows us to rewrite the field equations of the system as
\begin{equation}
  \nabla_{[\mu} \mathcal{S}_{\nu]\gamma} = 0,
  \label{eq:feqs-codazzi-form}
\end{equation}
where
\begin{equation}
  \mathcal{S}_{\nu\gamma} = \frac{\mathcal{R}_{\nu\gamma} - C \, \partial_{\nu} \phi \partial_\gamma \phi }{\mathcal{V}(\phi)}.
  \label{eq:our-codazzi}
\end{equation}

The equation \eqref{eq:feqs-codazzi-form} indicates that the tensor \(\mathcal{S}\) is a Codazzi tensor \cite{besse07_einst}.\footnote{In some references the quantities satisfying this kind of relations are said to be \emph{in equilibrium} at each point, see for example Sec. III.5 of Ref. \cite{schouten13_ricci}.} In the purely gravitational scenario, the equivalent of Eq. \eqref{eq:feqs-codazzi-form} (\(\nabla_{[\mu} \mathcal{R}_{\nu]\gamma} = 0\)) is a well-known generalisation of Einstein's equations, which is equivalent (through the differential Bianchi identity) to the condition of harmonic curvature, i.e. \(\nabla_{\lambda} \ri{\mu\nu}{\lambda}{\rho} = 0\) \cite{derdzinski80_class_certain_compac_rieman_manif,derdzinski82_compac_rieman_manif_with_harmon_curvat,derdzinski85_rieman}.

Following the strategy from Ref. \cite{castillo-felisola18_cosmol}, we shall consider three types of solutions to the problem posed by Eq. \eqref{eq:feqs-codazzi-form}, to know: (i) \(\mathcal{S}\) vanishes; (ii) \(\mathcal{S}\) is a parallel tensor; or (iii) \(\mathcal{S}\) is a Codazzi tensor.

Restricting ourselves to isotropic and homogeneous affinely connected spaces, the ansatz for the symmetric connection is given by \cite{castillo-felisola18_beyond_einstein}\footnote{Greek indices take values over the whole of the spacetime coordinates (\(\mu,\nu = 0, 1, 2, 3\)), while Latin indices are valued only on the space coordinates (\(i,j = 1, 2, 3\)).}
\begin{equation}
  \label{eq:ansatz-connection}
  \begin{aligned}
    \Gamma_{t}{}^{t}{}_{t} & = f(t), & \Gamma_{i}{}^{t}{}_{j} & = g(t) S_{i j}, \\
    \Gamma_{t}{}^{i}{}_{j} & = h(t) \delta^{i}_{j}, & \Gamma_{i}{}^{j}{}_{k} & = \gamma_{i}{}^{j}{}_{k},
  \end{aligned}
\end{equation}
where
\begin{align}
  S_{ij} =
  \begin{pmatrix} 
    \frac{1}{1 - \kappa r^2} & 0 & 0 \\
    0 & r^2 & 0 \\
    0 & 0 & r^2 \sin^2\theta 
  \end{pmatrix},
\end{align}
and
\begin{equation}
  \begin{aligned}
    \gamma_{r}{}^{r}{}_{r} & = \frac{\kappa r}{1 - \kappa r^2},
    &
    \gamma_{\theta}{}^{r}{}_{\theta} & = - r \left(1 - \kappa r^2 \right),
    \\
    \gamma_{\varphi}{}^{r}{}_{\varphi} & = - r \left(1 - \kappa r^2 \right) \sin^2 \theta,
    &
    \gamma_{r}{}^{\theta}{}_{\theta} & = \frac{1}{r},
    \\
    \gamma_{\varphi}{}^{\theta}{}_{\varphi} & = - \cos \theta \sin \theta,
    &
    \gamma_{r}{}^{\varphi}{}_{\varphi} & =\frac{1}{r},
    \\
    \gamma_{\theta}{}^{\varphi}{}_{\varphi} & = \frac{\cos \theta}{\sin \theta}.
    & &
  \end{aligned}
\end{equation}
The parameter \(\kappa\) might take the values \(1\), \(0\) or \(-1\). By choosing an adequate parametrisation of the \(t\)-coordinate, the \(\ct{t}{t}{t}\) component of the connection can be set to zero, i.e. \(f = 0\) \cite{castillo-felisola21_aspec_polyn_affin_model_gravit_three}.\footnote{It is worth to highlight that the Levi-Civita connection derived from the Friedmann--Robertson--Walker metric, parameterised with a \emph{scale factor} \(a\), is compatible with the ansatz in Eq. \eqref{eq:ansatz-connection} through the identifications: \(f = 0\), \(g = a \dot{a}\) and \(h = \frac{\dot{a}}{a}\).} In addition, the scalar field depends just on the time coordinate,
\begin{equation}
  \phi = \phi(t).
  \label{eq:ansatz-phi}
\end{equation}

From the cosmological ansatz, it follows that the components of the Ricci tensor are
\begin{align}
  \mathcal{R}_{tt} & = - 3 ( \dot{h} + h^2 ),
  &
    \mathcal{R}_{ij} & = ( \dot{g} + g h + 2 \kappa ) S_{ij},
\end{align}
while the \(\mathcal{S}\) tensor differs from it solely in the \(tt\) component,
\begin{align}
  \label{eq:S-tensor-components}
  \mathcal{S}_{tt} & = - 3 ( \dot{h} + h^2 ) - C (\dot{\phi})^2,
  &
    \mathcal{S}_{ij} & = ( \dot{g} + g h + 2 \kappa ) S_{ij},
\end{align}

\section{Overview of solutions}
\label{sec:solutions}
In a previous article \cite{castillo-felisola18_cosmol} (see also its corrigendum \cite{castillo-felisola23_corrig}), we analysed the solutions of the purely gravitational sector of the polynomial affine model of gravity. Before scanning the solutions of the polynomial affine gravity coupled to the scalar field, we present a brief review of the results in the purely gravitational sector, however, as mentioned above, we use the \emph{gauge fixing} \(f = 0\).

\subsection{Review of vacuum solutions}
\label{sec:purely-pag}
\subsubsection{Vanishing Ricci tensor}
\label{sec:vanishing-R}
In this case the field equations yield two differential equations,
namely 
\begin{align}
  \dot{h} + h^2 & = 0, & \dot{g} + gh + 2\kappa & = 0.
\end{align}
These can be solved exactly, giving
\begin{align}
  h(t) & = \frac{1}{t - t_0}, & g(t) & = - \frac{g_0 - t \kappa (t - 2 t_0)}{t - t_0}.
  \label{eq:affine-solution-ricci-0}
\end{align}

Under the assumption that the connection is a metric connection, i.e.
\(h(t) = \frac{\dot{a}}{a}\) and \(g(t) = \dot{a}a\), the equations are
\begin{align}
  \frac{\ddot{a}}{a} & = 0, & \ddot{a}a + 2\left(\kappa + \dot{a}^2\right) & = 0,
\end{align}
whose solutions is
\begin{align}
  a(t) & = t\sqrt{-\kappa} + a_0,
\end{align}
yielding
\begin{align*}
   h(t) & = \frac{\sqrt{-\kappa}}{t\sqrt{-\kappa} + a_0}, & g(t) & = \sqrt{-\kappa}\left(t\sqrt{-\kappa} + a_0\right).
\end{align*}
Notice that under the above assumptions, we have restrictions for the
geometry factor \(\kappa\), meaning that it can only take values as \(0\) and
\(-1\) to have a real geometry, whereas in the pure affine geometry,
there is no constraint for \(\kappa\).

\subsubsection{Parallel Ricci tensor}
\label{sec:parallel-R}
The parallel Ricci equation yields three differential equations, to know
\begin{align}
  \label{eq:parallel-ricci-hdd}
  \ddot{h} + 2h\dot{h} = \frac{\mathrm{d}}{\mathrm{d}t} \left( \dot{h} + h^2 \right) & = 0, \\
  2gh^2 - 2\kappa h - h\dot{g} + 3g\dot{h} & = 0, \\
  2gh^2 + 4\kappa h + h\dot{g} - g\dot{h} - \ddot{g} & = 0.
\end{align}
Since Eq. \eqref{eq:parallel-ricci-hdd} is a total derivative, its first integral yields a constant that we denote by \(c_0\).

The system of differential equations is solved analytically by the functions
\begin{align}
  \label{eq:affine-solution-parallel-ricci-h}
  h(t)
  & =
    \begin{cases}
      \sqrt{c_0}\tanh\left(\sqrt{c_0}(t - t_0)\right) & c_0 \neq 0
      \\
      h_{0} & c_0 = h_0^2
      \\
      \frac{1}{t - t_0} & c_0 = 0
    \end{cases},
  \\
  \label{eq:affine-solution-parallel-ricci-g}
    g(t) & =
           \begin{cases}
             \dfrac{\kappa \sinh{2 \sqrt{c_0} (t - t_0)}}{2\sqrt{c_0}} & c_0 \neq 0
             \\
             g_0 e^{2 h_0 t} + \frac{\kappa}{h_{0}} & c_0 = h_0^2
             \\
             - \dfrac{g_0 - \kappa t (t - 2 t_0)}{t - t_0} & c_0 = 0
           \end{cases}.
\end{align}
Notice that the solution is well-behaved for all values of \(c_0 \in \mathbb{R}\), because for negative values of \(c_{0}\) the hyperbolic functions get transformed to trigonometric functions, while the case with \(c_0 = 0\) simplifies to the vanishing Ricci scenario whose solution coincide with the reported in Eq. \eqref{eq:affine-solution-ricci-0}.

In the cases where \(c_0 \neq 0\) and \(\kappa \neq 0\), the Ricci tensor is symmetric and non-degenerated and might be used as a metric. In addition, given that it is parallel such metric is \emph{compatible} with the connection.

When one considers the connection descendent from the Friedmann--Robertson--Walker metric, the equations for the scale factor are
\begin{equation}
  \label{eq:a-parallel-ricci}
  \begin{aligned}
    \frac{\mathrm{d}}{\mathrm{d}t} \left( \frac{\ddot{a}}{a} \right) & = 0,
    \\
      \kappa + \dot{a}^2 - a\ddot{a} & = 0,
    \\
      \frac{4\dot{a}\left(\kappa + \dot{a}^2\right)}{a} - 3\dot{a}\ddot{a} - a\dddot{a} & = 0,
  \end{aligned}
\end{equation}
whose solution is
\begin{align}
  a(t) = \pm \frac{\sqrt{-\kappa}}{c_1} \sinh\left(c_1 \left(t + c_2 \right)\right),
  \label{eq:a-sol-parallel-ricci}
\end{align}
where the parameter \(c_1\) is given by the square root of the first integration constant of the first relation in Eq. \eqref{eq:a-parallel-ricci}, and thus it might be either a real or pure imaginary number. In addition, the only real and nontrivial scale factor requires that \(\kappa = -1\), however, it should be highlighted that in General Relativity all the vacuum Friedmann cosmologies are obtained for \(\kappa = -1\), see Table \ref{tab:summary-vacuum-friedmann-cosmologies}.

\begin{table}[htbp]
\caption{\label{tab:summary-vacuum-friedmann-cosmologies}Classification of vacuum Friedmann cosmologies in General Relativity \cite{dray14_differ}. The quantity \(q\) is defined as \(q = \sqrt{|\Lambda|/3}\).}
\centering
\begin{tabular}{c|c|c|c}
 & \(\Lambda < 0\) & \(\Lambda = 0\) & \(\Lambda > 0\)\\[0pt]
\hline
\(\kappa = 1\) & no solution & no solution & de Sitter\\[0pt]
 &  &  & \(a = \cosh(q t) / q\)\\[0pt]
\hline
\(\kappa = 0\) & no solution & Minkowski & de Sitter\\[0pt]
 &  & \(a = 1\) & \(a = \exp(q t)\)\\[0pt]
\hline
\(\kappa = -1\) & anti de Sitter & Minkowski & de Sitter\\[0pt]
 & \(a = \sin(q t) / q\) & \(a = t\) & \(a = \sinh(q t) / q\)\\[0pt]
\end{tabular}
\end{table}

From Eq. \eqref{eq:a-sol-parallel-ricci} one obtains the three vacuum Friedmann cosmological model of General Relativity as follows,
\begin{equation}
  a(t) =
  \begin{cases}
    \dfrac{\sinh(c_1 t)}{c_1} & \kappa = -1 \land c_1 \in \mathbb{R},
    \\[10pt]
    \dfrac{\sin(c_1 t)}{c_1} & \kappa = -1 \land c_1 \in \imath \mathbb{R},
    \\[10pt]
    t                       & \kappa = -1 \land c_1 = 0,
    \\
    0                        & \kappa = 0.
  \end{cases}
  \label{eq:a-sol-parallel-ricci-improved}
\end{equation}

\subsubsection{Harmonic curvature}
\label{sec:codazzi-R}
In the general case, the field equations yield a single differential equation,
\begin{align}
  4gh^2 + 2\kappa h + 2g\dot{h} - \ddot{g} = 0,
 \label{eq:harmonic-curvature}
\end{align}
for two undetermined functions. Although one can not find analytically expressions for \(h(t)\) and \(g(t)\) functions, it is possible to find families of solutions parametrised by one of such functions, e.g for a given function \(h\) we can solve the Eq. \eqref{eq:harmonic-curvature} for the function \(g\). It is worth to remark that the \emph{proper} solutions of the Eq. \eqref{eq:harmonic-curvature} have to exclude that cases analysed in Sec. \ref{sec:vanishing-R} and \ref{sec:parallel-R}.

From the definition of the components of the connection, \(\partial_{\mu} \vec{e}_{\rho} = \Gamma_{\mu}{}^{\lambda}{}_{\rho} \vec{e}_{\lambda}\), the function \(h\) determines the \emph{scaling} of the spatial base vectors as one moves along the \(t\)-coordinate. Hence, it indeed plays the role of an \emph{affine} Hubble parameter. Based on that premise, let us consider a few particular cases: (i) when the function \(h\) is constant, which in principle can be thought as an affine version of the de~Sitter space; (ii) when \(h\) grows linearly with \(t\); (iii) when \(h\) is a sine function of \(t\); and (iv) when \(h\) is an hyperbolic sine of the coordinate \(t\). The last two cases are inspired in the solutions for the \(h\) function obtained in Sec. \ref{sec:parallel-R}.

\begin{figure}[htbp]
\centering
\includegraphics[width=.9\linewidth]{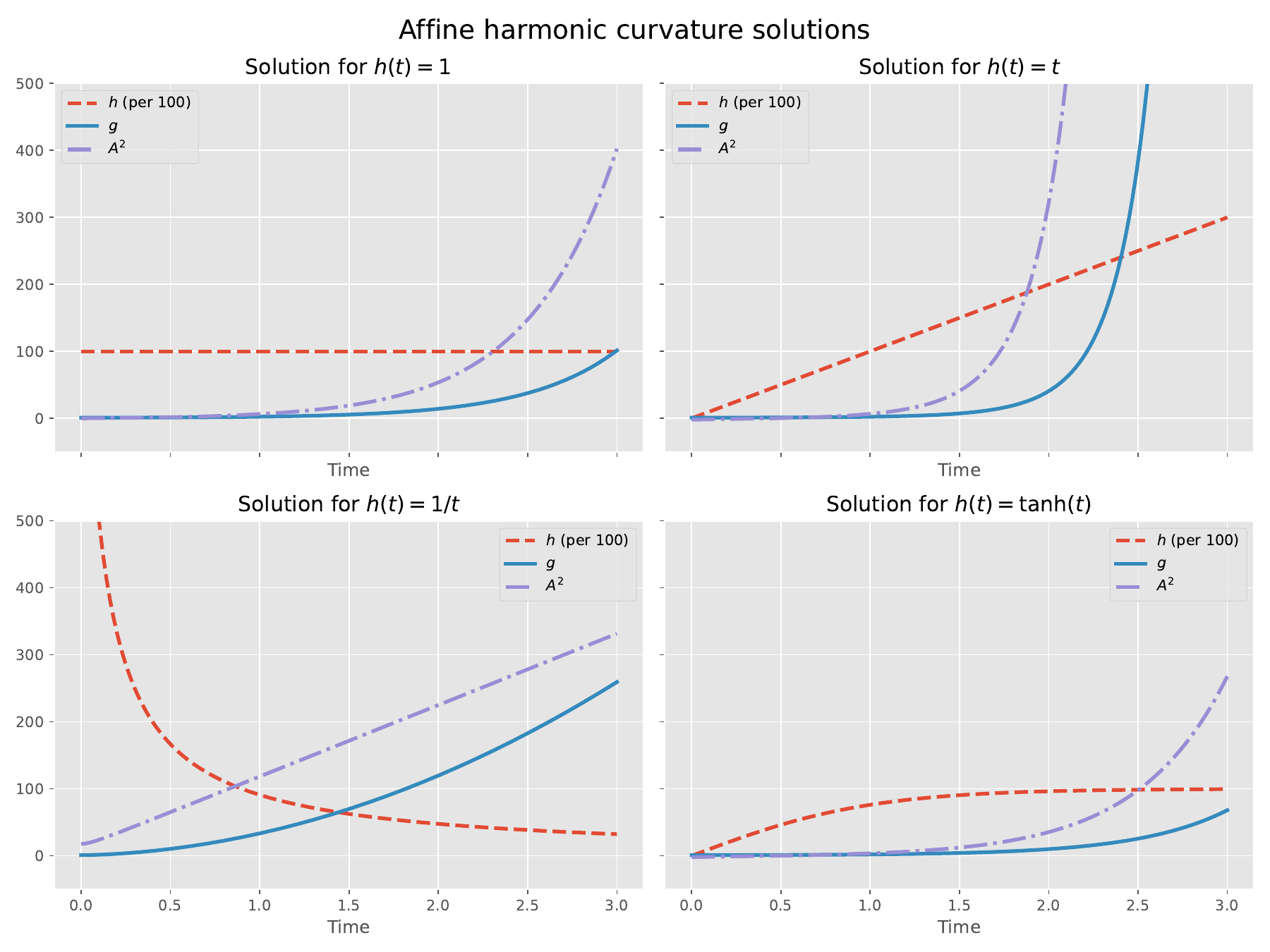}
\caption{\label{fig:harmonic-curvature-h-combined}Plot showing the behaviour of the functions \(g\) and \(h\). The function \(h\) has been taken as given, and the function \(g\) solves the Eq. \eqref{eq:harmonic-curvature} for the chosen function \(h\). The plots of the function \(h\) has been scaled by a factor of \(100\) to appear visible at the same scale than \(g\). The function \(A^2\) defines the value of the spatial components of the Ricci, which would be the square of the \emph{scale factor} if the Ricci tensor field is non-degenerated.}
\end{figure}

In most cases, the Eq. \eqref{eq:harmonic-curvature} can not be solved analytically, but one can still get some numerical results. The plots of the functions \(g\) and \(h\) for the cases described above are shown in Fig.~\ref{fig:harmonic-curvature-h-combined}.

\begin{figure}[htbp]
\centering
\includegraphics[width=.9\linewidth]{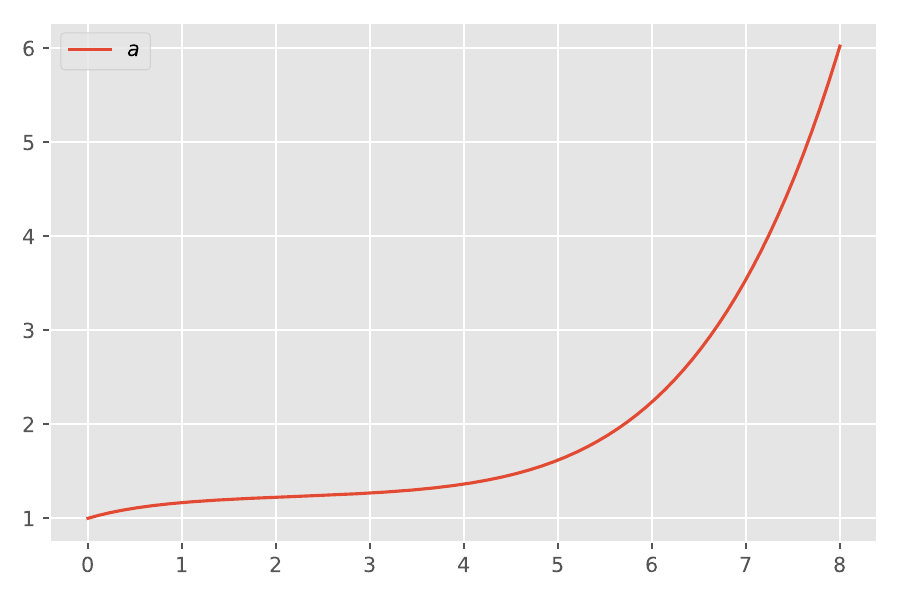}
\caption{\label{fig:harmonic-curvature-a}Plot of the scale factor \(a(t)\) with solves the Eq. \eqref{eq:harmonic-curvature-a}. The initial conditions used in the plot are \(a(0) = 1\), \(\dot{a}(0) = 0.3\) and \(\ddot{a}(0) = -0.43\), with \(\kappa = 1\).}
\end{figure}

As in the previous cases, under the assumptions that the connection is the Levi-Civita connection associated to a Friedmann--Robertson--Walker metric, the field equation is then
\begin{equation}
  \frac{2\dot{a}\left( \kappa + \dot{a}^2 \right)}{a} - \dot{a}\ddot{a} - a\dddot{a} = 0,
  \label{eq:harmonic-curvature-a}
\end{equation}
which can be solved numerically, see for example Fig.~\ref{fig:harmonic-curvature-a}.

\subsection{Solutions coupled with the scalar field}
\label{sec:coupled-pag}
\subsubsection{Vanishing \(\mathcal{S}\)}
\label{sec:vanishing-S}
Due to the symmetry of the scalar field, it follows that the Ricci tensor is degenerated (all their spatial components vanish) but nonvanishing. Hence, it does not define an \emph{emergent} metric on the manifold \cite{castillo-felisola20_emerg_metric_geodes_analy_cosmol}. The scalar potential, \(\mathcal{V}(\phi)\), does not affect the field equations,

The field equations can be read directly from Eq.~\eqref{eq:S-tensor-components}, 
\begin{equation}
  \begin{aligned}
    - 3 ( \dot{h} + h^2 ) - C \, (\dot{\phi})^2 & = 0,
    \\
      \dot{g} + g h + 2 \kappa & = 0.
  \end{aligned}
  \label{eq:vanishing-s}
\end{equation}
These are solved with ease in terms of the \(h\)-function,
\begin{align}
  \phi(t) & = \phi_0 \pm \sqrt{ - \frac{3}{C} } \int \mathrm{d}t \, \sqrt{\dot{h} + h^2},
  \label{eq:sol-phi-S0}
  \\
  g(t) & = e^{- \int \mathrm{d}t \, h} \left( g_0 - 2 \kappa \int \mathrm{d}t \, e^{\int \mathrm{d}t \, h} \right).
  \label{eq:sol-g-S0}
\end{align}
Note that Eq. \eqref{eq:sol-phi-S0} yields real solutions when the
constant \(C\) is negative, i.e. when the signs of the coupling
constants \(C_1\) and \(\alpha\) from Eqs. \eqref{eq:new-action} and
\eqref{eq:inv_metr_dens} are opposite.

\begin{figure}[htbp]
\centering
\includegraphics[width=.9\linewidth]{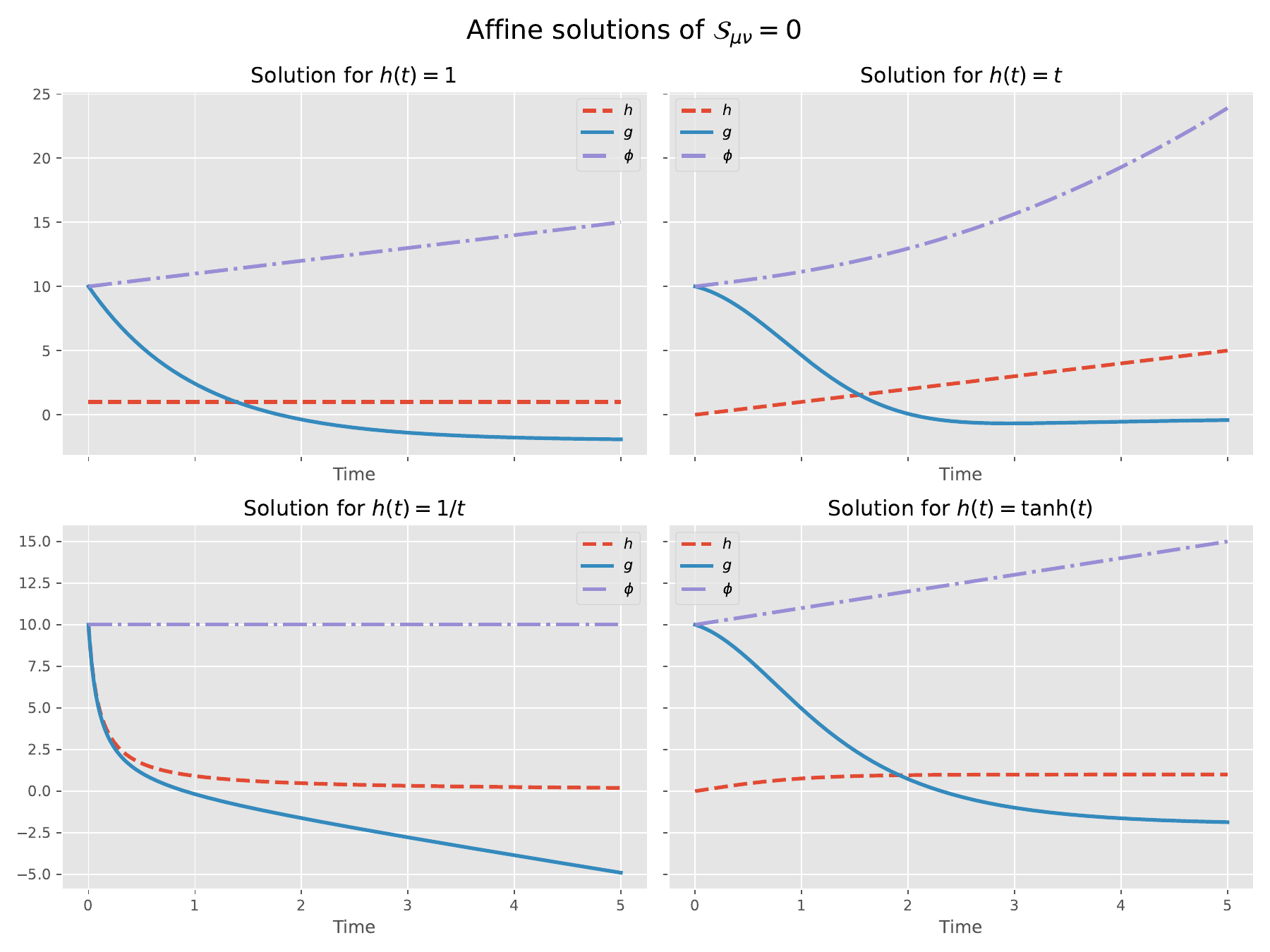}
\caption{\label{fig:vanishing-s}Numerical solutions for the Eqs. \eqref{eq:vanishing-s} for certain functions \(h\).}
\end{figure}

In previous articles we have argued that the \(h\)-function in our
model plays a similar role to that of the Hubble parameter on the
standard cosmology
\cite{castillo-felisola18_cosmol,castillo-felisola20_emerg_metric_geodes_analy_cosmol,castillo-felisola21_aspec_polyn_affin_model_gravit_three,castillo-felisola22_polyn_affin_model_gravit}.
Therefore, in the situation where \(h\) is a constant, \(h(t) = h_0\),
which would generalise the solution of (Anti) de Sitter, the Eqs.
\eqref{eq:sol-phi-S0} and \eqref{eq:sol-g-S0} reduce to
\begin{equation}
  \phi(t) = \phi_0 \pm \sqrt{- \frac{3}{C}} h_0 t \quad \text{and} \quad g(t) = g_0 e^{- h_0 t} - \frac{2 \kappa}{h_0}.
  \label{eq:sol-phi-g-S0-Ch}
\end{equation}
In principle the Eqs. \eqref{eq:sol-phi-S0} and \eqref{eq:sol-g-S0} can be
solved for a generic parametric function \(h\), however, we choose to
omit it since the function \(g\) is defined through the inverse of the
hypergeometric function \({}_2 F_1\) of \(h\).\footnote{We encourage the interested reader to check the existence of the general solution.} In Fig. \ref{fig:vanishing-s} the solutions \eqref{eq:sol-phi-g-S0-Ch} are plotted for a selection of the function \(h\).

\begin{figure}[htbp]
\centering
\includegraphics[width=.9\linewidth]{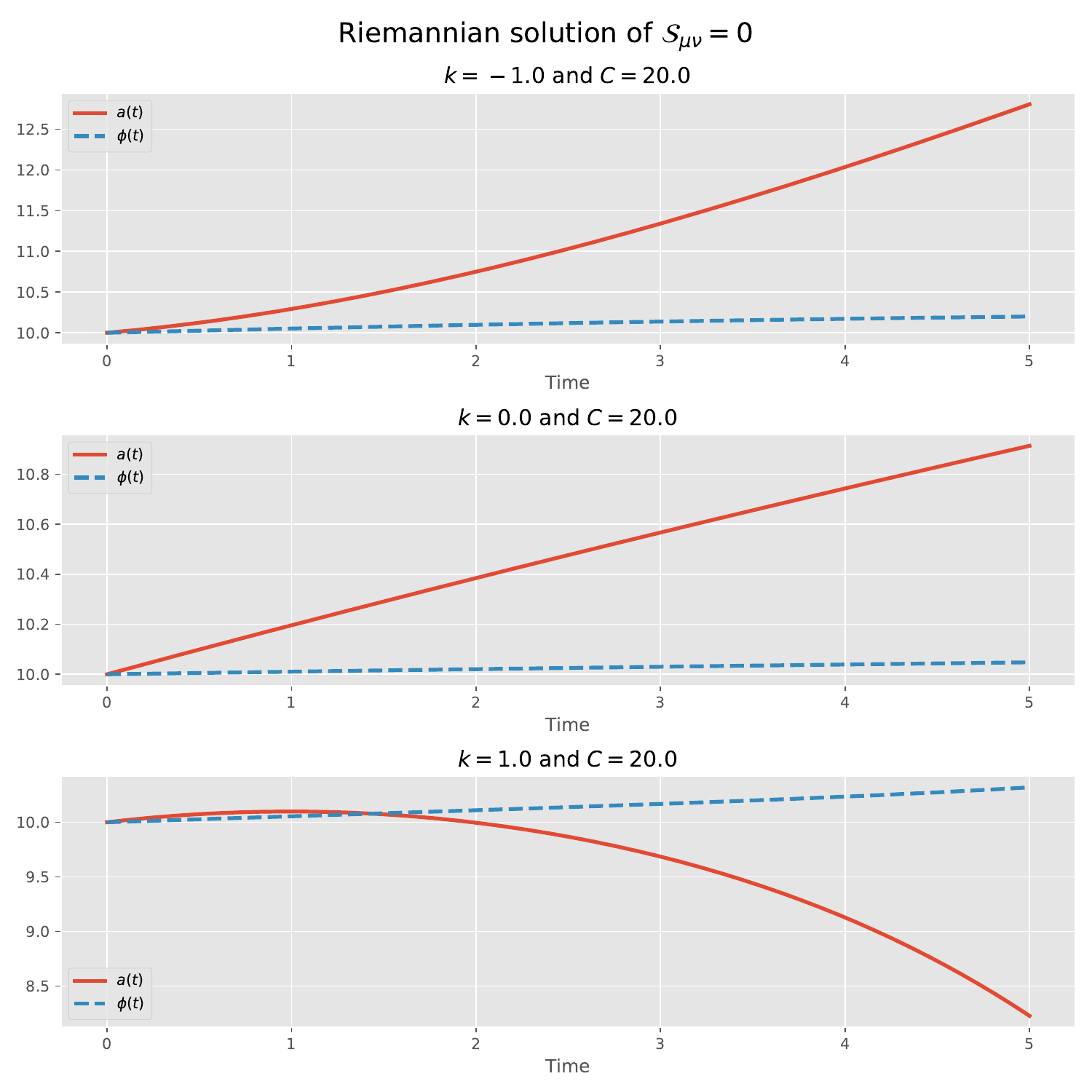}
\caption{\label{fig:vanishing-s-a}Numerical solution of the Eq. \eqref{eq:vanishing-s-a} for the three possible values of the constant \(\kappa\). The initial conditions have been set as \(a(0) = \phi(0) = 10.0\), and \(\dot{a}(0) = 0.2\).}
\end{figure}

The Riemannian equivalent of the above system is driven by the equations
\begin{equation}
  \begin{aligned}
    \dot{\phi} & = \sqrt{\frac{6}{C} \frac{\dot{a}^2 + \kappa}{a^2}},
    &
    \ddot{a} & = - 2 \frac{\dot{a}^2 + \kappa}{a^2},
  \end{aligned}
  \label{eq:vanishing-s-a}
\end{equation}
for these relations to be well-defined the constant \(C\) has to be strictly positive. The solution of these equations is shown in Fig. \ref{fig:vanishing-s-a}.

\subsubsection{Parallel \(\mathcal{S}\)}
\label{sec:parallel-S}
As shown in Appendix \ref{app:parallel-tensor}, in order for the tensor \(\mathcal{S}\) to be parallel the functions \(g\) and \(h\) have to be obtained from a potential function \(a\), turning this problem equivalent to the (pseudo-)Riemannian case, i.e. to General Relativity minimally coupled to a self-interacting massless scalar field. If we denote by \(g_{\mu\nu}\) the symmetric and non-degenerated parallel \(\binom{0}{2}\)-tensor with respect to the affine connection, the parallel condition for the tensor \(\mathcal{S}\) can be rewritten as
\begin{equation}
  \mathcal{R}_{\mu\nu} - C \, \partial_{\mu}\phi \partial_{\nu} \phi = \mathcal{V}(\phi) \, \Sigma \, g_{\mu\nu},
  \label{eq:parallel-S-g}
\end{equation}
where \(\Sigma\) is an integration constant. Equivalently, the condition can be expressed in the Einstein form,
\begin{equation}
  \mathcal{R}_{\mu\nu} - \frac{1}{2} \mathcal{R} g_{\mu\nu} = C \left( \partial_{\mu}\phi \partial_{\nu} \phi - \frac{1}{2} g_{\mu\nu} (\partial \phi)^2 \right) - \Sigma \, \mathcal{V}(\phi) \, g_{\mu\nu}.
  \label{eq:parallel-S-einstein}
\end{equation}
Note that for \(\Sigma = C/2\) the Eq. \eqref{eq:parallel-S-einstein} is exactly what is obtained in General Relativity.\footnote{The fixing of the constant \(\Sigma\) is achieved by a scaling of the original coupling constants in the potential \(\mathcal{V}(\phi)\).}

The above result is a generalisation of the one obtained in Ref. \cite{castillo-felisola18_einst_gravit_from_polyn_affin_model}, which would not include the self-interacting potential. As in the aforementioned reference, the field equation for the scalar field can be derived through the Bianchi identity (following the prescription suggested in Ref. \cite{bekenstein15_is_princ_least_action}) or by taking the divergence of Eq. \eqref{eq:parallel-S-einstein}, which after some simplifications yields
\begin{equation}
  C \, \nabla^{\mu} \nabla_{\mu} \phi - \Sigma \, \mathcal{V}^{\prime}(\phi) = 0,
  \label{eq:klein-gordon}
\end{equation}
where the prime denotes derivation with respect to the scalar field \(\phi\).

\subsubsection{\(\mathcal{S}\) is a Codazzi}
\label{sec:codazzi-S}
In the case of \(\mathcal{S}\) to be a Codazzi tensor, there is a single nontrivial equation,
\begin{dmath}
  C \mathcal{V}(\phi) g \dot{\phi}^2
  + \mathcal{V}(\phi) (4 g h^2 + 2 \kappa h + 2 g \dot{h} - \ddot{g})
  + \mathcal{V}'(\phi) \dot{\phi} (g h + 2 \kappa + \dot{g}) = 0,
  \label{eq:codazzi-S}
\end{dmath}
but this involves three unknown functions for a choice of potential \(\mathcal{V}(\phi)\), such functions are \(h(t)\), \(g(t)\) and \(\phi(t)\). Therefore, we cannot find analytic solutions without further conditions.

If we restrict ourselves to the Levi-Civita connection compatible with a metric of Friedmann--Robertson--Walker the system still requires complementary conditions, since the field equation will be determined by two functions, \(a(t)\) and \(\phi(t)\), for a certain potential \(\mathcal{V}(\phi)\).

Since the above equations are not enough to determine a solution of the set up, we follow two strategies to complement the Eq. \eqref{eq:codazzi-S}: (i) consider that the Ricci tensor field is parallel, and (ii) considering that the tensor \(\mathcal{S}\) possesses the form of a Friedmann--Robertson--Walker-like metric.\footnote{This method is similar to the one used to find solutions with harmonic curvature in Ref. \cite{castillo-felisola20_emerg_metric_geodes_analy_cosmol}.}

\subsubsection*{Covariantly constant Ricci tensor}
\label{sec:codazzi-S-parallel-R}
Noticing that the tensor \(\mathcal{S}\) differs from the Ricci by terms depending on the scalar field, if the Ricci tensor is parallel then the Eq. \eqref{eq:codazzi-S} turns into a equation for the scalar field \(\phi\).

For the sake of simplicity we consider from Eqs. \eqref{eq:affine-solution-parallel-ricci-h} and \eqref{eq:affine-solution-parallel-ricci-g}, the solution parameterised by \(c_0 = h_0^2\), and choose the case with \(\kappa = 0\) to avoid the degeneracy of the Ricci tensor. Under this considerations the affine connection is parameterised by the functions
\begin{equation*}
  h(t) = h_0 \text{ and } g(t) = g_0 e^{2 h_0 t},
\end{equation*}
yielding a Ricci tensor of the form
\begin{align*}
  \mathcal{R}_{00} & = - 3 h_0^2, & \mathcal{R}_{ij} & = 3 g_0 h_0 e^{2 h_0 t} S_{ij}.
\end{align*}
Finally the condition of \(\mathcal{S}\) to be a Codazzi tensor turns into
\begin{equation*}
  C \mathcal{V}(\phi) \phi^2 + 3 h_0 \mathcal{V}'(\phi) \dot{\phi} = 0,
\end{equation*}
or equivalently
\begin{equation}
  \dot{\phi} + \frac{C}{3 h_0} \left(\frac{\mathcal{V}(\phi)}{\mathcal{V}'(\phi)}\right) \phi^2 = 0.
  \label{eq:codazzi-S-parallel-R-phi}
\end{equation}
Note that in the above equations we have omitted the trivial solutions \(\dot{\phi} = 0\), because all the effect of the presence of the scalar field vanish, and the set up turns to be equivalent to the purely gravitational model.

The above equation is, as expected, an equation for the scalar field for a given inflationary potential. In the following we shall consider three choices of potentials.

Considering a power-law type of potential \(\mathcal{V}(\phi) = \alpha \phi^n\) the Eq. \eqref{eq:codazzi-S-parallel-R-phi}
\begin{equation}
  \dot{\phi} + \frac{3 n h_0}{C \phi} = 0,
\end{equation}
whose solution is 
\begin{equation}
  \phi(t) = \pm \sqrt{2 \phi_0 - \frac{6 n h_0 t}{C}}.
  \label{eq:phi-solution-power-law-c1}
\end{equation}
Interestingly, the behaviour of the potential in Eq. \eqref{eq:phi-solution-power-law-c1} indicates (assuming the constants \(\phi_0\), \(C\), \(n\) and \(h_0\) as positive) that after some finite period of time the scalar field will vanish, providing a mechanism for ending the inflationary epoch.

Considering the potential of quintessence, \(\mathcal{V}(\phi) = \alpha e^{- \beta \phi}\), the equation for the scalar field is
\begin{equation}
  \dot{\phi} - \frac{3 \beta h_0}{C} = 0,
\end{equation}
whose solution is given by
\begin{equation}
  \phi(t) = \phi_0 + \frac{3 \beta h_0 t}{C}.
  \label{eq:phi-solution-quintessence-c1}
\end{equation}

Finally, considering the Starobinsky potential, \(\mathcal{V}(\phi) = \alpha \left( 1 - e^{- \beta \phi} \right)^2\), the scalar field satisfies the field equation
\begin{equation*}
   C \left(e^{\beta  \phi} - 1\right) \dot{\phi} + 6 \beta  h_0  = 0,
\end{equation*}
whose nontrivial solution is expressed in terms of the Lambert \(W\)-function as follows,\footnote{The Lambert \(W\) function takes a number \(z\) and returns the number \(w\) such that the \(z = w \, e^w\).}
\begin{equation}
  \phi(t) = - C_1 - \frac{1}{\beta} W_0 \left( - e^{\beta \left( \frac{6 \beta h_0}{C} t - C_1 \right)} \right) + \frac{6 \beta h_0}{C} t.
  \label{eq:phi-solution-starobinsky-c1}
\end{equation}
In the last equation \(C_1\) is the integration constant.

\subsubsection*{Ansatz for the \(\mathcal{S}\) tensor}
\label{sec:codazzi-S-ansatz}
Based in the ansatz for isotropic and homogeneous covariant symmetric \(\binom{0}{2}\)-tensors, we search for solutions to the field equation \eqref{eq:codazzi-S} with the prescription that \(\mathcal{S}\) is characterised by
\begin{equation}
  \mathcal{S}_{tt} = - \mathcal{S}_{0}, \quad \mathcal{S}_{ij} = A(t) \, S_{ij},
  \label{eq:codazzi-S-ansatz}
\end{equation}
with \(\mathcal{S}_0\) a constant.

Replacing the ansatz into Eq. \eqref{eq:codazzi-S}, one obtains the relation
\begin{equation}
  \dot{A} - h \, A - g \, \mathcal{S}_0 = 0,
  \label{eq:codazzi-S-relation}
\end{equation}
together with the \emph{compatibility} conditions
\begin{align}
  \label{eq:codazzi-S-compt1}
  3 \left(\dot{h} + h^2 \right) + C \dot{\phi}^2 & = S_0 \mathcal{V}(\phi),
  \\
  \label{eq:codazzi-S-compt2}
  \dot{g} + g h & = A \mathcal{V}(\phi).
\end{align}
This set of equations can be solve in terms of one of the unknowns, for a given potential.

Based in the solutions obtained in previous cases, we shall consider the simplest scenario, obtained when we restrict ourselves to the constant \(h\) case, i.e. \(h = h_0\).

Using the power-law potential, \(\mathcal{V}(\phi) = \alpha \, \phi^n\), the Eq. \eqref{eq:codazzi-S-compt1} turns into
\begin{equation}
  {\dot{\phi}}^2 - \frac{\alpha \mathcal{S}_{0}}{C} \phi^n + \frac{3 h_{0}^2}{C} = 0.
  \label{eq:codazzi-S-phi-power-law}
\end{equation}
Equation \eqref{eq:codazzi-S-phi-power-law} can be formally solved for generic values of the power \(n\),\footnote{The solution involves the inverse of the hypergeometric function \({}_2 F_1\)} and for the lower values the  scalar field solving the field equation is
\begin{equation}
  \phi(t) =
  \begin{cases}
    \phi_0 \pm t \sqrt{\frac{\alpha \mathcal{S}_0 - 3 h_0^2}{C}} & n = 0 \\[5pt]
    \frac{3 h_0^2}{\alpha \mathcal{S}_{0}} + \frac{\alpha \mathcal{S}_0}{4 C} \left( t - t_0 \right)^2 & n = 1 \\[5pt]
    \sqrt{- \frac{3 h_0^2}{\alpha \mathcal{S}_0}} \sinh \left( \sqrt{\frac{\alpha \mathcal{S}_{0}}{C}} (t - t_0) \right) & n = 2
  \end{cases}.
  \label{eq:codazzi-S-phi-power-law-solution}
\end{equation}

Finally, from the Eqs. \eqref{eq:codazzi-S-relation} and \eqref{eq:codazzi-S-compt2} one obtains an ordinary differential equation for the function \(A\),
\begin{equation}
  \ddot{A} - A \left( h_0^2 + \mathcal{S}_0 \mathcal{V}(\phi) \right) = 0.
  \label{eq:codazzi-S-eqn-A}
\end{equation}
Since the equation for \(A\) depends on the function \(\phi\) through the potential, the solutions get complicated as the potential gets a richer structure. Based in the solutions in Eq. \eqref{eq:codazzi-S-phi-power-law-solution}, for the constant potential the function \(A\) is given by
\begin{equation}
  A(t) =
    A_0 \, e^{- t \sqrt{h_0^2 + \alpha \mathcal{S}_0}} + A_1 \, e^{+ t \sqrt{h_0^2 + \alpha \mathcal{S}_0}},
  \label{eq:codazzi-S-A-power-law-solution}
\end{equation}
while for the linear potential the function \(A\) is determined in terms of the Weber's parabolic cylinder functions, \(D_p(t)\).\footnote{The parabolic cylinder functions are the \emph{special functions} that solve the parabolic part of the Laplace equation, much like what the Bessel functions are to the radial equation of the Laplacian in cylindrical coordinates.}

For the case of the quadratic potential we were unable to integrate the Eq. \eqref{eq:codazzi-S-eqn-A} analytically, so if the reader is interested in the quadratic potential scenario the goal should be (in our opinion) pursue using numerical methods.

\section{Discussion of results}
\label{sec:discussion}
In this article we extend the purely Polynomial Affine Model of Gravity in four dimensions by coupling it to a self-interacting scalar field. We focus our attention to the torsion-free sector of the model, and inquire the cosmological scenarios compatible with our assumptions.

As mentioned in the body of the article, the field equations of the model take a very simple form when one consider the torsion-free sector (with or without the scalar field), reminding us the Codazzi equation---and also the definition if the Cotton tensor \cite{cotton99_sur,garcia04_cotton_tensor_rieman_spacet} in the absence of a metric tensor---.

The Codazzi equation for a tensor field \(T\) (see Eq. \eqref{eq:feqs-codazzi-form}) possesses three types of solutions: (i) when the tensor itself vanishes; (ii) when the tensor is parallel, and; (iii) when the tensor is properly a Codazzi tensor field. Evidently, the space of solutions of the first type, \(\mathcal{M}_{(i)}(T)\), is a subspace of the space of solutions of type (ii), \(\mathcal{M}_{(ii)}(T)\), and the latter is a subspace of the space of solutions of type (iii), \(\mathcal{M}_{(iii)(T)}\), i.e.
\begin{equation}
  \mathcal{M}_{(i)}(T) \subseteq \mathcal{M}_{(ii)}(T) \subseteq \mathcal{M}_{(iii)}(T).
  \label{eq:solution-spaces}
\end{equation}

In the article we found proper cosmological solutions of \(\mathcal{M}_{(iii)}\) and \(\mathcal{M}_{(ii)}\), that is solutions that lie in the spaces \(\mathcal{M}_{(iii)}/\mathcal{M}_{(ii)}\) and \(\mathcal{M}_{(ii)}/\mathcal{M}_{(i)}\) respectively (for both  \(\mathcal{R}_{\mu\nu}\) and \(\mathcal{S}_{\mu\nu}\)). Hence, we shown that in Eq. \eqref{eq:solution-spaces} there is a strict ordering of spaces.

Firstly, we review the vacuum cosmological solutions of the polynomial affine model of gravity. This analysis represents a simplification of the results reported in Ref. \cite{castillo-felisola20_emerg_metric_geodes_analy_cosmol}, obtained after re-parameterise the \emph{time} coordinate (see Appendix B of Ref. \cite{castillo-felisola21_aspec_polyn_affin_model_gravit_three}). Our current study provides a different insight into the subject, since we consider the existence of connection-descendent (also named emergent) metrics provided the affine solutions.

Given that the (symmetric component of the) Ricci tensor field is the \emph{a priori} candidate of emergent metric, the affine solutions from \(\mathcal{M}_{(i)}(\mathcal{R})\) does not posses a \emph{natural} emergent metric. Nonetheless, if one assumes that the affine connection is the Levi-Civita connection of a yet-to-be-known Friedmann--Robertson--Walker-like metric, the solution is a re-parametrization of the Minkowski geometry (\(a(t) = t\)). Interestingly, the affine solution in Eq. \eqref{eq:affine-solution-ricci-0} cannot be in general be associated to a Riemannian geometry (apparently the integrability condition requires the constant \(g_{0}\) to vanish).

The case discussed above drives us toward the ``conditions on a connection to be a metric connection'' \cite{schmidt73_condit_connec_to_be_metric_connec}, which leads us to the notion of holonomy group of a connection. Unfortunately, most of the advances in the subject of holonomy lay on the existence of a metric, or even constraint to metrics with Euclidean signature. For example, the Berger's classification of irreducible holonomy groups is based on the ground of Riemannian geometry \cite{berger55_sur_les_group_dholon,joyce00_compac_manif_special_holon,gross03_calab_yau_manif_relat_geomet}, but a Lorentzian equivalent is not yet known (see for example Ref. \cite{galaev08_holon_loren}).

It is worth mentioning that in affine models (such as Polynomial Affine Gravity) if it is possible to define connection-descendent metrics, in principle there is no \emph{a priori} determined signature for such metric. This flexibility in the signature, allows affine solutions that do not fit into the premises of General Relativity. It can be seen in the fact that the cosmological affine solutions do not constrain the values of the \(\kappa\) parameter.

We found that the space of cosmological affine solutions  \(\mathcal{M}_{(ii)}(\mathcal{R})\) extends extensively the space of cosmological vacuum solutions of General Relativity, and particularly the cosmological constant appears as an integration constant (like in the case of Unimodular Gravity \cite{ng91_unimod_theor_gravit_cosmol_const,bengochea23_clarif_prevail_miscon_unimod_gravit}). However, when one restricts to the subspace of solutions where the connection is derived from a Friedmann--Robertson--Walker-like metric, the subspace \(\mathcal{M}_{(ii)}\left(\mathcal{R}\big|_{FRW}\right)\) coincides with the vacuum solutions of General Relativity. Like in the previous scenario, the affine solutions do not restrict the values of the parameter \(\kappa\).

In the case of the Ricci as Codazzi tensor (which is equivalent to the harmonic curvature), there is a single nontrivial component of the field equations, depending on two functions \(g\) and \(h\). Therefore, the strategy is to solve the field equation on one variable in terms of the other. Nonetheless, when one restricts to a Levi-Civita connection, that depends on a single function (the scale factor \(a\)), it is possible to solve numerically the field equation \eqref{eq:harmonic-curvature-a}.

The numerical solutions presented in Figs. \ref{fig:harmonic-curvature-h-combined} and \ref{fig:harmonic-curvature-a} show the behaviour of some proper solutions on \(\mathcal{M}_{(iii)}(\mathcal{R})\), in particular the latter figure shows the richness in the evolution depending of the value of the initial conditions. It is worth mentioning that in both figures it is possible to explain the accelerated expansion of the Universe.

Then, we turn our attention to the scenario of Polynomial Affine Gravity coupled with a scalar field. 

When the case with vanishing \(\mathcal{S}_{\mu\nu}\) is considered, i.e. solutions on the space \(\mathcal{M}_{(i)}(\mathcal{S})\), one notices that the scalar field couples to the gravitational sector through the \((t,t)\)-component of the tensor \(\mathcal{S}\), and also that the self-interacting potential of the scalar field plays no role in the dynamics of the system. Hence, since there are two independent field equations but three undetermined functions (\(g\), \(h\) and \(\phi\)), it is possible to solve the dynamics of the system in terms of one of the unknowns.

It has been argued in the presiding sections that the geometrical meaning of the component of the connection \(\Gamma_{t}{}^{i}{}_{j}\), allows us to interpret the function \(h\) as a sort of Hubble parameter, that tells us how a spatial vector changes along its direction,\footnote{Only along its direction because this component of the connection is proportional to a Kronecker delta in the indices \(i\) and \(j\).} as it is moved forward in the coordinate \(t\). For this reason we parameterise the solutions with \(h\). A case of interest is that were \(h = h_0\) is constant, yielding the affine notion of (Anti) de Sitter and Minkowski spaces with an homogeneous scalar field.

We found that the space of solutions \(\mathcal{M}_{(ii)}(\mathcal{S})\), thanks to the results shown in Appendix \ref{app:parallel-tensor}, is related to the space of solutions of the system Einstein gravity coupled with a free massless scalar field. In particular, if one choose \(\Sigma = C/2\) in Eq. \eqref{eq:parallel-S-einstein}, the system is equivalent to Einstein's gravity (without cosmological constant) coupled with the scalar field.

Interestingly, the parallel case allows to obtain a field equation for the scalar field from the differential Bianchi identity, which turns to be a Klein--Gordon equation \eqref{eq:klein-gordon}.

Note that for \(\Sigma \neq C/2\) the field equations \eqref{eq:parallel-S-einstein} yield solutions that, in general, do not have equivalent in General Relativity coupled with a scalar field.

It is worth mentioning that, as expected from a gravitational system coupled with a self-interacting scalar field, even if the dynamics of the scalar field is not affected by a shift of the potential (\(V(\phi) \to V(\phi) + V_0\)), see Eq. \eqref{eq:klein-gordon}, such shift modifies the \emph{Einstein} equation \eqref{eq:parallel-S-einstein}. The modification induced by the aforementioned shift is interpreted as a modification in the value of the cosmological constant, where \(\Lambda = \Sigma \, V_0\).

The solutions in \(\mathcal{M}_{(iii)}(\mathcal{S})\) are (in general terms) more relevant than the previous ones, in the sense that they depart further from the known solutions in General Relativity. Moreover, is in this scenario that the scalar potential plays a main role in the plot. Since the space of solutions \(\mathcal{M}_{(iii)}(\mathcal{S})\) is defined by a sole equation that depends on three independent unknowns, it is highly degenerated. In order to be able of finding solutions, we followed two approaches: (i) Considering the simplification where the Ricci tensor field is parallel , and (ii) Proposing an ansatz for the \(\mathcal{S}\)-tensor, and ask for self-consistency to solve the field equation.

In the first approach, the Ricci tensor is an affine generalisation of either flat or (Anti) de Sitter, so the condition that \(\mathcal{S}\) is a Codazzi tensor turns into an equation for the dynamics of the (self-interacting) scalar field. That equation can be solved once the self-interacting potential has been chosen.

On the other hand, the self-consistency of second approach introduces new equations, Eqs. \eqref{eq:codazzi-S-compt1} and \eqref{eq:codazzi-S-compt2}, enhancing the set of differential equations. The new system of allows to solve the three unknowns without further assumptions, once the self-interacting potential of the scalar field has been chosen. Interestingly, even in the simplest solution for the power-law potential (see Eq. \eqref{eq:codazzi-S-A-power-law-solution}) might account for a rich behaviour of the scale factor \(A\), depending on the signs and modulus of the constants \(A_1\) and \(A_2\).

In view of our findings, the (vacuum) cosmological solutions of the polynomial affine model of gravity posses a degree of complexity that in General Relativity can only be attributed to material effects, providing a best suited arena to provide a geometrical interpretation of phenomenology that in the \(\Lambda\)CDM is due to the dark sector.

We want to highlight that very recently there is a growing interest in the cosmological effects derived from the affine connection. There are a few works that we would like to highlight. In Ref. \cite{salvio23_inflat_reheat_throug_indep}, the author proposes a model of inflation and reheating that is driven by a geometrical object that encodes the additional degrees of freedom introduced when the connection is not a Riemannian connection. However, his approach differs from ours since it possess a metric, i.e. it is a metric-affine model. In Ref. \cite{aoki23_cosmol_pertur_theor_metric,heisenberg23_cosmol_telep_pertur,heisenberg23_gauge_invar_cosmol_pertur}, the authors independently propose formalisms to analyse the cosmological perturbations in metric-affine theories of gravity.\footnote{A different approach to analyse the cosmological perturbations in affine models is currently been developed \cite{castillo-felisola23}.} These new tools will allow us to extract further information from the models that would be compared with the cosmological observations.


\begin{acknowledgements}
  The authors are grateful to N. Zambra, J. Vaca, M. Morocho, A. Zerwekh and Ivan Schmidt\footnote{OCF wants to dedicate this article to the memory of Ivan, who has passed away recently.} for their fruitful discussions and invaluable support during the development of this research.

  We are indebted to the authors and countless contributors of the free and open software community, for maintaining the projects \texttt{SAGEmath}, \texttt{sagemanifolds} and \texttt{cadabra} \cite{stein18_sage_mathem_softw_version,gourgoulhon18_sagem_version,peeters07_symbol_field_theor_with_cadab,peeters07_introd_cadab,peeters07_cadab,brewin19_using_cadab_tensor_comput_gener_relat,kulyabov19_new_featur_secon_version_cadab}, which has been used extensively to complete this work.
\end{acknowledgements}

\appendix

\section{Isotropic and homogeneous parallel \((0,2)\)-tensors}
\label{app:parallel-tensor}
In Sec. \ref{sec:parallel-R} and \ref{sec:parallel-S} we are interested in finding isotropic and homogeneous tensors of type \((0,2)\), which are parallel with respect to the connection in Eq. \eqref{eq:ansatz-connection}.

Since the isotropic and homogeneous \((0,2)\)-tensor is symmetric and characterised by two functions of the \(t\)-coordinate (see for example Ref. \cite{castillo-felisola18_cosmol}), the condition of parallelism
\begin{equation}
  \nabla_{\lambda} T_{\mu\nu} = \partial_{\lambda} T_{\mu\nu} - \Gamma_{\lambda}{}^{\sigma}{}_{\mu} T_{\sigma\nu} - \Gamma_{\lambda}{}^{\sigma}{}_{\nu} T_{\mu\sigma} = 0,
\end{equation}
has only four nontrivial components,
\begin{align}
  \label{eq:parallel-000}
  & \lambda = \mu = \nu = 0  \colon & & \partial_{t} T_{tt} = 0, \\
  \label{eq:parallel-0ii}
  & \lambda = 0; \; \mu = \nu = i \colon & & \partial_{t} T_{ii} - 2 h T_{ii} = 0, \\
  \label{eq:parallel-ijj}
  & \lambda = i; \; \mu = \nu = j \colon & & \partial_i T_{jj} - 2 \gamma_{i}{}^{j}{}_{j} T_{jj} = 0, \\
  \label{eq:parallel-i0j}
  & \lambda = i; \; \mu = 0; \; \nu = j \colon & & - \Gamma_{i}{}^{j}{}_{t} T_{jj} - \Gamma_{i}{}^{t }{}_{j} T_{tt} = 0.
\end{align}
From Eq. \eqref{eq:parallel-000} it follows that \(T_{tt} = - b_0\) is constant. Equation \eqref{eq:parallel-ijj} is equivalent to the parallelism of the three-dimensional metric \(s_{ij}\), and therefore
\begin{equation}
  T_{ij} = \Xi(t) \, s_{ij}.
  \label{eq:parallel-tii}
\end{equation}
Then, substituting the above into Eqs. \eqref{eq:parallel-0ii} and \eqref{eq:parallel-i0j}, we obtain respectively that
\begin{equation}
  \Xi(t) \propto \exp \left( 2 \int \mathrm{d}t \, h \right),
  \quad \text{ and } \quad 
  \Xi(t) = \frac{b_0 g}{h},
  \label{eq:solution-Xi}
\end{equation}
which are compatible if both \(h\) and \(g\) are obtainable from a \emph{potential} function \(a\),
\begin{align}
  g & = a \dot{a}, & h & = \frac{\dot{a}}{a}, & \Xi & \propto a^2.
  \label{eq:solution-Xi-a}
\end{align}
Hence, the connection is equivalent to the Riemannian connection and the parallel tensor is a Friedmann--Robertson--Walker-like metric.

\end{document}